# Joint Chance-Constrained Economic Dispatch Involving Joint Optimization of Frequency-related Inverter Control and Regulation Reserve Allocation

Ye Tian, *Student Member, IEEE*, Zhengshuo Li, *Senior Member, IEEE,* Wenchuan Wu, *Fellow, IEEE,* Miao Fan, *Senior Member, IEEE*

*Abstract*—The issues of uncertainty and frequency security could become significantly serious in power systems with the high penetration of volatile inverter-based renewables (IBRs). These issues make it necessary to consider the uncertainty and frequency-related constraints in the economic dispatch (ED) programs. However, existing ED studies rarely proactively optimize the control parameters of inverter-based resources related to fast regulation (e.g., virtual inertia and droop coefficients) in cooperation with other dispatchable resources to improve the system frequency security and dispatch reliability. This paper first proposes a joint chance-constrained economic dispatch model that jointly optimizes the frequency-related inverter control, the system up/down reserves, and base-point power for the minimal total operational cost. In the proposed model, multiple dispatchable resources including thermal units, dispatchable IBRs and energy storage are considered, and the (virtual) inertias, the regulation reserve allocations and the base-point power are coordinated. To ensure the system reliability, the joint chance-constraint formulation is also adopted. Additionally, since the traditional sample average approximation (SAA) method cost much computational burden, a novel mix-SAA (MSAA) method is proposed to transform the original intractable model into a linear model that can be efficiently solved via commercial solvers. The case studies validated the satisfactory efficacy of the proposed ED model and the efficiency of the MSAA.

## NOMENCLATURE

*Indices and sets*
- $b$  Index of bus from 1 to $Nb$
- $e$  Index of energy storage (ES) from 1 to $Ne$
- $g$  Index of thermal unit from 1 to $Ng$
- $i$  Index of scenario from 1 to $N$
- $l$  Index of transmission line from 1 to $Nl$
- $w$  Index of dispatchable inverter-based renewable (DIBR) from 1 to $Nw$
- $E$  Set of ESs
- $G$  Set of thermal units
- $G_b, G_b^c$  Set of thermal units and auto generation control (AGC) units at bus $b$
- $W_b, E_b$  Set of DIBRs and ESs at bus $b$
- $W$  Set of DIBRs
- $N$  Set of bus nodes
- $[n]$  Set of sampling scenarios

This work was supported by the National Natural Science Foundation of China under Grant 52007105. Y. Tian and Z. Li are with the School of Electrical Engineering, Shandong University, Jinan 250061, China. Zhengshuo Li is the corresponding author (e-mail: zsli@sdu.edu.cn).
W. Wu is with the State Key Laboratory of Power Systems, Department of Electrical Engineering, Tsinghua University, Beijing 100084, China (e-mail: wuwench@tsinghua.edu.cn).
Miao Fan is with Siemens Industry, Inc., Schenectady, NY 12305, US (email: fanmiao@ieee.org).

*Parameters*
- $C_g(\cdot)$  Fuel cost function of thermal unit $g$
- $c_g^\uparrow, c_g^\downarrow$  Cost coefficients of the up/down reserve of thermal unit $g$
- $c_g^r$  Cost coefficient of redispatch power of thermal unit $g$
- $c_e^{E\uparrow}, c_e^{E\downarrow}$  Cost coefficients of the up/down reserve of ES $e$
- $c_e^{loss}$  Cost coefficients of charging/discharging loss of ES $e$
- $c_w^W$  Cost coefficients of curtailment of DIBR $w$
- $D_O$  Damping coefficient of the whole system
- $\overline{D}_k$  Upper bound of droop coefficient of DIBR or ES $k$
- $d_b, \tilde{d}_b$  Forecast/actual load at bus $b$
- $\overline{E}_e, \underline{E}_e$  Max/min limit of energy of ES $e$
- $\eta_{ch}, \eta_{dis}$  Coefficient of charging/discharging efficiency of ES $e$
- $F_l$  Power capacity of transmission line $l$
- $f_0$  Normal frequency
- $\overline{H}_k$  Upper bound of virtual inertia of DIBR or ES $k$
- $\hbar_b, \tilde{\hbar}_b$  Forecast/actual power of uncontrollable IBR at bus $b$
- $\overline{p}_g, \underline{p}_g$  Max/min limit of power output of thermal unit $g$
- $\overline{p}_e^E$  Maximum limit of charging/discharging power of ES $e$
- $\overline{p}_w^W$  Forecasted maximum output of DIBR $w$
- $\overline{\tilde{p}}_w^W$  Actual maximum output of DIBR $w$
- $p_i$  Probability of sampling scenario $i$
- $R_k^u, R_k^d$  Up/down reserve of DIBR or ES $k$
- $R_G, R_g$  Droop coefficient of the aggregated thermal units/thermal unit $g$
- $\overline{r}_{p,g}^u, \overline{r}_{p,g}^d$  Maximum up/down ramping rate of thermal unit $g$
- $S_{l,b}^F$  Power transfer distribution factor from bus $b$ to line $l$
- $\Delta t$  Dispatch time period
- $\Delta f$  Frequency variation of the system
- $\Delta \overline{f}_{rate}$  Maximum acceptable absolute value of rate of change of frequency (RoCoF)
- $\Delta \overline{f}_{max}$  Maximum acceptable absolute value of the maximum frequency deviation
- $\Delta \overline{f}_{ss}$  Maximum acceptable absolute value of the steady-state frequency deviation
- $\Delta p_L$  Power disturbance of the whole system
- $\delta_F, \delta_L$  Significance level of dynamic frequency violation and power flow overload on transmission line
- $\delta_{DIBR}, \delta_{SFR}$  Significance level of insufficient up reserves of DIBR and SFR reserves

*Variables*
- $D_w$  Droop control coefficient of DIBR $w$
- $D_e$  Droop control coefficient of ES $e$
- $H_w$  Virtual inertia of DIBR $w$
- $H_e$  Virtual inertia of DIBR $e$
- $p_w^W$  Base-point power of DIBR $w$
- $p_e^E$  Charging/discharging power of ES $e$
- $p_e^{loos}$  Power loss of ES $e$

| $p_g$ | Base-point power of thermal unit $g$ |
| $r_g^\uparrow, r_g^\downarrow$ | Up/down reserve capacity of thermal unit $g$ |
| $r_e^{E\uparrow}, r_e^{E\downarrow}$ | Up/down reserve capacity of ES $e$ |
| $\alpha_g$ | Adjustable AGC allocation factor of thermal unit $g$ |
| $z_i$ | Binary indicator variable for scenario $i$ |

## I. INTRODUCTION

INVERTER-based renewables (IBRs) have developed rapidly. As reported in [1], the share of renewable energy in the US has grown to over 24% of total power sources and will increase to 44% in 2050, and the proportion for China is nearly 31% [2]. Renewable generation not only causes the issues of intermittency and uncertainty but also raises the concerns about the frequency security of power system operations [3]. In particular, due to the reduced system inertia [4], the maximum *rate of change of frequency* (RoCoF) and the maximum frequency deviation are being increasingly and widely considered [5][6].

Economic dispatch (ED) programs are applied to optimally determine the base-point power and the up/down reserves of multiple dispatchable power resources, e.g., thermal units, dispatchable IBRs (DIBRs), and energy storage (ES). In the past few years, a variety of robust and stochastic programming approaches have been introduced into ED studies to handle the challenge of uncertainty [7][8]. As frequency security issues are becoming prominent, recent works [3]-[6], [9]-[16] suggest that ED problems should also involve frequency-related constraints to ensure the frequency security in the dispatch time period. Ref. [9] considered dynamic frequency characteristics, [10][11] improved the frequency regulation requirement estimation method to construct associated constraints, and [12][13] suggested incorporating dynamic secondary frequency regulation (SFR) constraints. Furthermore, as it becomes technically feasible to adjust inverters' control parameters (e.g., droop coefficients) on the ED time scale, incorporating dispatchable inverter-based resources (e.g., DIBRs and ESs) in ED programs to leverage their potential for providing virtual inertia response (IR) and fast regulating services, e.g., primary/fast frequency regulation (PFR), has attracted much attention [14]. In [15], the virtual inertia provided by wind farms is modeled as a random variable to determine the unit commitment and regulation reserves. However, in the above studies, jointly optimizing the inverter-based resources' control parameters **(e.g., virtual inertia and droop coefficients)** related to fast regulation together with other system dispatchable resources are not taken into account, and some methods may be inaccurate or computationally prohibitively expensive [16].

Recently, [17] studied the optimal allocation of virtual inertia and droop control coefficients of DIBRs and ESs, but the method did not consider the high-frequency related constraints, which may cause potential operation risks, which can be demonstrated by the example in Fig. 1. At midday, the net load (load minus distributed generation) becomes low, and power disturbances could cause both up and down overlimit issues of the system frequency. If the units were dispatched as suggested by [17], which neglects the high-frequency related constraints, the system operator would have insufficient headroom for downward regulation. In addition, the second drawback of the method in [17] is that it modeled the constraints involving random variables as individual chance constraints (ICCs). However, it has been reported that ICCs "*do not give strong guarantees on the feasibility probability of the entire system*" [18]. On the other hand, joint chance constraints (JCCs) can overcome the drawback of ICCs and have a better safety guarantee [19].

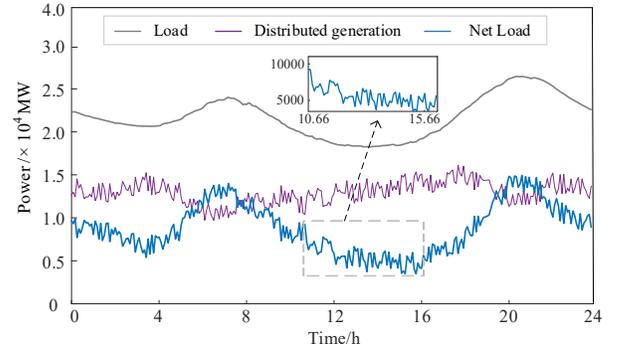

**Fig. 1.** Power trajectory of load and distributed generation.

In summary, the above review of the state-of-the-art reveals that the following factors should be involved in the ED problem. ***First***, besides the consideration of the base-point power and reserves, the frequency-related IR and the PFR control parameters of DIBRs and ESs should also be optimized, rather than regarding them as constants, for desirable frequency control behaviors in the dispatch time period. ***Second***, JCCs should be adopted to model the operational constraints considering the impact of uncertainty, e.g., frequency-related constraints, both up and down reserve constraints, transmission line flow constraints, etc. In particular, the high-frequency related constraints omitted by [15][17][20] should be carefully considered to avoid potential risks, as explained above. ***Finally***, the automatic generation control (AGC) allocation factors of thermal units should be optimized dynamically rather than taking them as constants [17]. This dynamic optimization strategy can significantly enhance the economical operation of power systems [21].

However, with all the above factors being considered, a joint chance-constrained ED model with a variety of complicated constraints and variables (e.g., controller parameters of DIBRs and ESs, reserve capacity, AGC allocation factors, and base-point power of multiple dispatchable resources) comes out. The JCC formulation usually makes the optimization model extremely challenging to solve [22]. Some popular approaches, e.g., the Bonferroni inequality approach [23] and sample average approximation (SAA) [24], have respective limitations. Specifically, the Bonferroni inequality approach could encounter difficulty in determining the confidence level of each ICC [23] and cause the suboptimality issue [25]. The SAA method that introduces binary indicator variables would yield a large-scale mixed integer linear programming (MILP) problem when a large number of sampling scenarios are needed to obtain effective empirical distribution information. A large-scale MILP problem costs a huge number of computational burden and can be unfeasible for real-time economic dispatch.

To overcome the aforementioned challenges, this paper proposes a joint chance-constrained economic dispatch (JCED) model that jointly optimizes frequency-related inverter control, system up/down reserves, and base-point power for the minimal total operational cost. Specifically, JCCs are adopted to model the operational constraints considering the uncertainty influence so that the base-point power and the up/down regulation reserves of thermal units, DIBRs and ESs are rationally allocated to achieve the economic optimal operation point of the system. The frequency-related inverter control parameters, i.e., virtual inertias and droop coefficients of DIBRs and ESs, are also optimized to meet the requirements of the RoCoF and the maximum frequency deviations. Furthermore, inspired by the techniques of the mixing inequality[1] and the aggregated mixing inequality [22], a novel mix-SAA (MSAA) method is proposed to transform JCED into a linear programming (LP) model that can be efficiently solved via commercial solvers. The major contributions of this paper are summarized as follows.

1) Compared with [17] and other works reviewed above, the proposed JCED model addresses **both high- and low-frequency issues** and adds the related constraints by proactively optimizing inverter-based resources' control parameters in cooperation with other system dispatchable resources. In addition, the complicated but reliable JCC formulations are adopted, and the AGC allocation factors are dynamically optimized to enhance the dispatch reliability and economy. As demonstrated in case studies, compared with the counterpart model in [17], the proposed JCED model materially improves the system security and economy against the uncertainty and frequency issues.

2) By applying the mixing inequality and aggregated mixing inequality, the MSAA method only takes approximately 10% of the computing time of traditional SAA in case studies with slightly compromising the optimization result accuracy. Therefore, MSAA prevents the potential issue of the large-scale MILP caused by using the traditional SAA and the difficulty of *a priori* determining the confidence level of each ICC in the Bonferroni inequality approach [23].

The rest of this paper is organized as follows. Section II introduces the frequency-related control of DIBRs and ESs, an equivalent model of the system frequency response and frequency-related constraints. Section III articulates the proposed JCED model. Section IV presents the MSAA method. Case studies and analysis are given in Section V. Section VI summarizes the paper and discusses future research.

## II. Frequency-related Constraints Considering DIBRs and ESs

*1) Frequency-related control of DIBRs and ESs*

A DIBR or an ES can provide virtual inertia and fast regulation responses to power systems. [9] shows that its control behavior can be modeled by the following equation (1).

$$\Delta p_k = -(2H_k \frac{d\Delta f}{dt} + D_k \Delta f), k \in \{W, E\} \quad (1)$$

---

[1] The effectiveness of the mixing inequality in dealing with simple chance-constrained unit commitment problems is preliminarily explored in[26], but it does only consider one ICC and not consider the JCCs. Our MSAA can handle the **problem involving multiple JCCs.**

Moreover, according to [14], it is technically feasible to adjust the control parameters $H_k, D_k$ in the ED time-scale for a desired purpose.

*2) Equivalent Model of the System Frequency Response*

There are usually tens of or even more DIBRs, ESs and thermal units in a power system. Therefore, instead of using the single control equation (1) and the dynamic model of each thermal unit, an equivalent model [27] of the system frequency response is utilized in (2) with the associated control block diagram shown in Fig. 2.

$$\Delta p_G^{eq} - \Delta p_L = 2(H_G^{eq} + H_W^{eq} + H_E^{eq})\frac{d\Delta f}{dt} + (D_O + D_W^{eq} + D_E^{eq})\Delta f \quad (2)$$

where $\Delta p_k$ represents the output variation of DIBR or ES $k$; $\Delta p_G^{eq}$ represents the output variation of the aggregated thermal unit, which is also a function of $\Delta f$ as shown in Fig. 2; $H_G^{eq}, H_W^{eq}, H_E^{eq}$ represents the inertia of the aggregated thermal unit and the virtual inertia of the aggregated DIBR and ES, respectively; and $D_W^{eq}, D_E^{eq}$ represents the droop coefficients of the aggregated DIBRs and ESs, respectively. The detailed derivation of (2) and the aggregated parameters in Fig. 2 are introduced in appendix A.

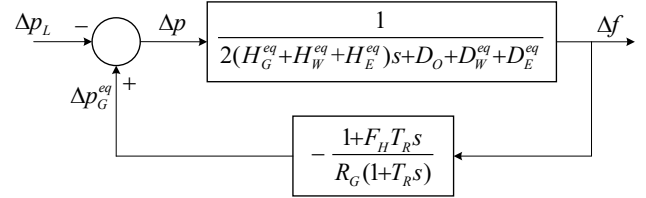

**Fig. 2.** Control block diagram of the equivalent model of the system frequency response[4].

*3) Frequency-related Constraints*

After obtaining the system frequency response model in (2), three widely used frequency-related indices [28], *maximum RoCoF*, *maximum frequency deviation*, and *steady-state frequency deviation* shown in (3a)-(3c) are applied in the subsequent ED model. The derivation of these frequency-related inequalities is shown in Appendix B.

**Maximum RoCoF:** $\quad |\frac{-\Delta p_L}{2(H_G^{eq} + H_W^{eq} + H_E^{eq})}| \leq \Delta \overline{f}_{rate} \quad (3a)$

**Maximum frequency deviation:** $\quad |\frac{-\Delta p_L}{F_{un}(H_W^{eq}, D_W^{eq}, H_E^{eq}, D_E^{eq})}| \leq \Delta \overline{f}_{max} \quad (3b)$

**Steady-state frequency deviation:** $\quad |\frac{-\Delta p_L}{D_O + D_W^{eq} + D_E^{eq} + R_G^{-1}}| \leq \Delta \overline{f}_{ss} \quad (3c)$

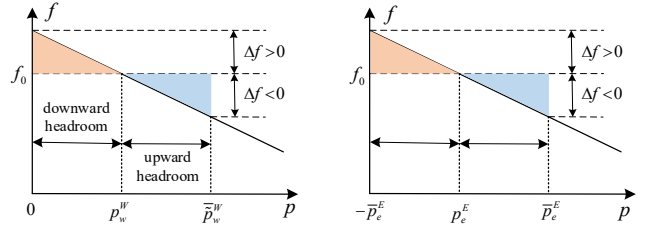

**Fig. 3.** Upward and downward regulation headroom for (a) DIBR, (b) ES.

Besides the frequency-related constraints (3), additional constraints should also be considered to ensure that DIBRs and ESs preserve sufficient headroom for the frequency regulation. As illustrated in Fig. 3, only the droop control with regard to $\Delta f$ is considered and represented by the oblique line. When $\Delta f < 0$, the power of the DIBR and ES should increase accordingly, which is fulfilled only if there is sufficient upward headroom. When $\Delta f > 0$, similar actions can be done except that no downward headroom is needed for the DIBRs which allow de-loading operation, i.e., proactively reducing its output whenever needed [9]. Therefore, the following constraints (4) that ensure sufficient frequency-related control headroom for DIBRs and ESs should also be involved in the subsequent ED model.

Upward headroom: $\quad R_k^u \geq |2H_k \frac{d\Delta f}{dt} + D_k \Delta f|, \Delta f > 0, k \in \{W, E\}$ (4a)

Downward headroom: $\quad R_k^d \geq |2H_k \frac{d\Delta f}{dt} + D_k \Delta f|, \Delta f > 0, k \in E$ (4b)

Constraint (4) is called as *the PFR constraints of DIBRs and ESs* in this paper, since it is similar to the PFR constraints of thermal units. However, (4) is much more complicated than the thermal units' PFR constraints, because $H_k, D_k$ are control variables, and $d\Delta f/dt$ and $\Delta f$ are also dependent on them, as implied in Fig. 2. In other words, (4) is a nonlinear constraint with $H_k, D_k$. Since being slightly conservative is acceptable by system operators in practice, a relaxed version (5) is derived to replace (4). The detailed derivation is shown in Appendix B.

$$\begin{aligned} R_k^u &\geq 2H_k \Delta \bar{f}_{rate} + D_k \Delta \bar{f}_{\max}, \quad \forall k \in \{W, E\} \\ R_k^d &\geq 2H_k \Delta \bar{f}_{rate} + D_k \Delta \bar{f}_{\max}, \quad \forall k \in E \end{aligned} \quad (5)$$

In summary, the frequency-related constraints (3) and (5) will be involved in the ED model established in the next section.

### III. PROPOSED JCED MODEL

#### A. Modeling assumptions

In this study, the following common modeling assumptions are adopted.

**1) Assumption I:** The regulation requirements of IR, PFR, and SFR are modeled individually.

**2) Assumption II**: Only the regulation requirements for the maximum power disturbance are considered.

**3) Assumption III**: DIBRs and ESs only participate in IR and PFR to prevent frequency change.

Assumptions I and II are widely adopted in the extant research [15][17]. For Assumption III, subject to the power grid guidelines and the control method of power electronic resources, there are two main ways for the participation of the DIBRs and ESs in the regulation [29]: the participation in PFR only [30] and the participation in both PFR and SFR [31]. This paper adopts Assumptions I, II and III (i.e., the first one of above two main ways) [15][17], which are widely accepted in optimal power dispatch considering the maximum RoCoF and the maximum frequency deviation. More information can be found in [15] [28]-[32]. In addition, it is worth noting that the proposed JCED model can be readily changed by making DIBRs and ESs assume a certain percentage of disturbance in the SFR constraints given in Section III. B, if the participation of DIBRs and ESs in SFR are necessary.

The modeling procedure of the proposed JCED are illustrated in Fig. 4. Thermal units provide inertia and reserves, while DIBRs and ESs provide virtual inertias and droop control support in the IR and PFR processes. The control parameters of DIBRs and ESs, reserve capacity, AGC allocation factors, and base-point power of multiple dispatchable resources are optimized together in the JCED model to meet the regulation requirements for all possible disturbances in the dispatch time period. Thus, the proposed JCED model can make more reasonable scheduling decisions.

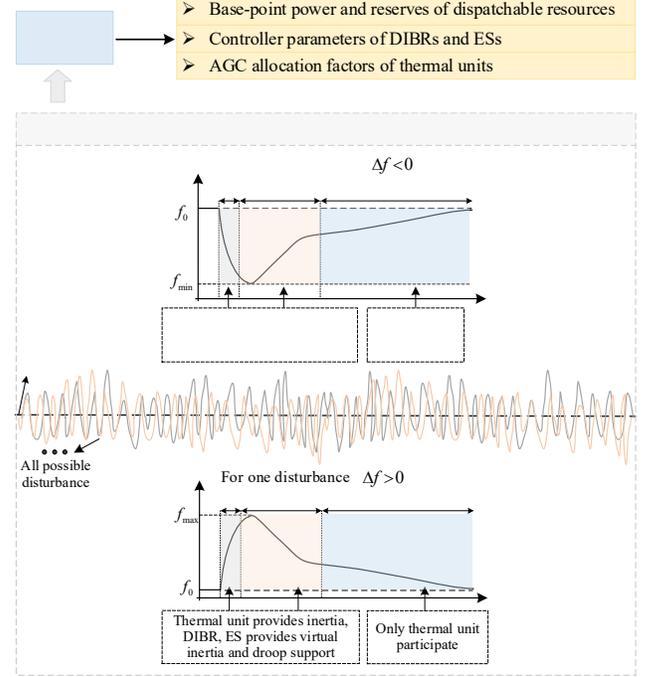

Fig. 4. Illustration of the modeling procedure of JCED.

#### B. Formulation

This section introduces the detailed formulation of the proposed JCED model, in which the decision variables include the inverter control parameters $H_w, H_e, D_w, D_e$, the AGC allocation factors $\alpha_g$, the up/down regulation reserve capacities $r_g^\uparrow, r_g^\downarrow$, $r_e^{E\uparrow}, r_e^{E\downarrow}$ and the base-point power $p_g, p_w^W, p_e^E$ of multiple dispatchable resources. Moreover, it should be noted that according to [33], the up/down reserve capacities of thermal units are defined for both PFR and SFR.

*1) Objective function*

$$\begin{aligned} \min &\sum_{g \in G} \{C_g(p_g) + c_g^\uparrow r_g^\uparrow + c_g^\downarrow r_g^\downarrow\} + \sum_{w \in W} \mathbb{E}_{\bar{p}_w^W}(c_w^W(\bar{p}_w^W - p_w^W)) \\ &+ \sum_{e \in E} \{c_e^{loss} p_e^{loos} + c_e^{E\uparrow} r_e^{E\uparrow} + c_e^{E\downarrow} r_e^{E\downarrow}\} \\ &+ \sum_{g \in G} \mathbb{E}_{\Delta p_L(\zeta)} c_g^r (\min\{r_g^+(\zeta), r_g^\uparrow\} + \min\{r_g^-(\zeta), r_g^\downarrow\}) \end{aligned} \quad (6)$$

The objective function includes the base-point power cost, the reserve capacity cost and the steady-state redispatch power cost of thermal units, the expected curtailment cost of DIBRs, the charging/discharging loss cost and the reserve capacity cost of ESs, where $\mathbb{E}_{\Delta p_L(\zeta)}, \mathbb{E}_{\bar{p}_w^W}$ are the expectation operators with respect to random variable $\Delta p_L(\zeta), \bar{p}_w^W$, respectively, and $r_g^+(\zeta), r_g^-(\zeta)$ are nonnegative auxiliary variables that satisfy $r_g^+(\zeta) - r_g^-(\zeta) = \alpha_g \Delta p_L(\zeta)$. Since the redispatch power cost in (6) is nonconvex, which makes the model intractable, an approximation of the redispatch power cost is given in Appendix C. It is worth noting that in accordance with [17][19], since only the frequency-related constraints (3) and (5) are involved instead of specific and complete frequency dynamic trajectories [12], only the reserve capacity cost and the steady-state redispatch power cost are considered in (6). The mileage cost [12] of the delivered reserves in the system dynamic process could be considered in the future work.

*2) Power balance equations under uncertain injections*

$$\sum_{g\in G} p_g + \sum_{w\in W} p_w^W + \sum_{e\in E} p_e^{ES} = \sum_{b\in N}(d_b - \hbar_b) \quad (7)$$

$$\tilde{d}_b = d_b + \zeta_b^d, \quad \tilde{\hbar}_b = \hbar_b + \zeta_b^\hbar \quad (8)$$

$$\sum_{b\in N}(\zeta_b^d - \zeta_b^\hbar) = \Delta p_L(\zeta) \quad (9)$$

Constraint (7) enforces the system power balance. Constraint (8) indicates the relationship between actual $\tilde{d}_b, \tilde{\hbar}_b$ and forecast $\tilde{d}_b, \tilde{\hbar}_b$, where the random variables $\zeta_b^d, \zeta_b^\hbar$ represent the corresponding uncertain forecast error [19]. Constraint (9) represents the disturbance caused by $\zeta_b^d, \zeta_b^\hbar$, where $\zeta$ is a simplified notation associated with $\zeta_b^d, \zeta_b^\hbar$, and $\Delta p_L(\zeta)$ can also be set as the disturbance caused by an uncertain fault.

*3) Generation constraints of thermal units and DIBRs*

$$p_g + r_g^\uparrow \le \overline{p}_g, \quad p_g - r_g^\downarrow \ge \underline{p}_g, \forall g \quad (10a)$$

$$r_g^\uparrow \le \overline{r}_{p,g}^u \Delta t, \quad r_g^\downarrow \le \overline{r}_{p,g}^d \Delta t, \forall g \quad (10b)$$

$$0 \le p_w^W \le \overline{p}_w^W, \forall w \quad (11)$$

Constraint (10a) ensures that the base-point power and the up/down regulation reserve capacities of thermal units are within capacity limits. Constraint (10b) ensures that the up/down reserves are within the ramping capacity limits in the dispatch time period. Constraint (11) enforces that the base-point power of DIBRs are within the maximum forecast output limits.

*4) ES constraints*

$$p_e^E + r_e^{E\uparrow} \le \overline{p}_e^E, \quad p_e^E - r_e^{E\downarrow} \ge -\overline{p}_e^E, \forall e \quad (12a)$$

$$\underline{E}_e \le E_0 - (p_e^E + p_e^{loss})\Delta t \le \overline{E}_e, \forall e \quad (12b)$$

$$p_e^{loss} = \max\{(1/\eta_{dis} - 1)p_e^E, (\eta_{ch} - 1)p_e^E\} \quad (12c)$$

$$p_e^{loss} \ge (1/\eta_{dis} - 1)p_e^E, p_e^{loss} \ge (\eta_{ch} - 1)p_e^E \quad (12d)$$

The ES models are from [17], including the base-point power and the reserve capacities constraints, the state of charge constraints and the charging/discharging loss constraints, respectively represented by (12a)-(12c), and the relaxation of (12c) is shown in (12d). The detailed derivation can be found in [17].

*5) Frequency-related constraints*

According to the frequency-related constraint (3), the linear maximum RoCoF constraint (ⅰ) and the steady-state frequency deviation constraint (ⅲ) can be readily obtained, and the linearized maximum frequency deviation constraints (ⅱ) are derived in Appendix B to avoid intractable nonlinearity. Since the constraints (ⅰ)(ⅱ)(ⅲ) jointly guarantee the dynamic frequency security of the system and $\Delta p_L(\zeta)$ is uncertain, the JCC formulation shown in (13a) is adopted, where $\mathbb{P}_{\Delta p_L(\zeta)}$ is the probability distribution with respect to the random variable $\Delta p_L(\zeta)$. The system operator can adapt confidence level $1-\delta_F$ according to his risk preference. The adjustable range of the virtual inertia and the droop coefficients of DIBR and ES is shown in (13b). Based on the analysis in Fig. 3, the up-reserve constraint of DIBR is modeled as a JCC (14). In contrast, based on the relaxed *PFR constraints* (5), the up/down PFR reserve capacities of ESs are constrained in (15), and similarly, the up/down PFR reserve capacities of thermal units are constrained as deterministic constraints in (16) as in [15].

$$\mathbb{P}_{\Delta p_L(\zeta)}\begin{cases} H_W^{eq} + H_E^{eq} \ge \frac{|\Delta p_L(\zeta)|}{2\Delta \overline{f}_{rate}} - H_G^{eq} & (\text{ⅰ}) \\ H_W^{eq} + H_E^{eq} \ge \alpha_m - \beta_m(D_W^{eq} + D_E^{eq}), \forall m & (\text{ⅱ}) \\ D_W^{eq} + D_E^{eq} \ge \frac{|-\Delta p_L(\zeta)|}{\Delta \overline{f}_{ss}} - (D_O + R_G^{-1}) & (\text{ⅲ}) \end{cases} \ge 1-\delta_F \quad (13a)$$

$$0 \le H_k \le \overline{H}_k, \quad 0 \le D_k \le \overline{D}_k, \quad k \in \{W, E\} \quad (13b)$$

$$\mathbb{P}_{\overline{p}_w^W}\{\overline{p}_w^W - p_w^W \ge 2H_w \Delta \overline{f}_{rate} + D_w \Delta \overline{f}_{\max}, \forall w\} \ge 1-\delta_{DIBR} \quad (14)$$

$$\begin{aligned} r_e^{E\uparrow} \ge 2H_j \Delta \overline{f}_{rate} + D_j \Delta \overline{f}_{\max} \\ r_e^{E\downarrow} \ge 2H_j \Delta \overline{f}_{rate} + D_j \Delta \overline{f}_{\max} \end{aligned} \quad (15)$$

$$r_g^\uparrow \ge \Delta \overline{f}_{ss}/R_g, \quad r_g^\downarrow \ge \Delta \overline{f}_{ss}/R_g \quad (16)$$

*6) SFR constraints*

SFR is designed to mitigate the system frequency deviation. The reserves of DIBRs and ESs only participate in PFR based on Assumption III. Therefore, the thermal units need sufficient SFR reserves to balance the system power in steady-state conditions. In this condition, dynamically optimizing the AGC allocation factors $\alpha_g$ will improve the dispatch economy [21] by influencing the redispatch power cost in (6). Additionally, to make safer SFR reserve decisions under uncertain net load disturbances, a complicated but reliable JCC formulation (17) is adopted, where $\alpha_g$ are optimized with the constraint (18).

$$\mathbb{P}_{\Delta p_L(\zeta)}\{-r_g^\downarrow \le \alpha_g \Delta p_L(\zeta) \le r_g^\uparrow, \forall g\} \ge 1-\delta_{SFR} \quad (17)$$

$$\sum_{g\in G} \alpha_g = 1 \quad (18)$$

*7) Transmission line power flow constraints*

$$\mathbb{P}_{\zeta_b^d,\zeta_b^\hbar}\begin{cases} |\sum_{b\in N} S_{l,b}^F\{\sum_{g\in G_b}(p_g + \alpha_g \Delta p_L(\zeta)) \\ + \sum_{w\in W_b} p_w^W + \sum_{e\in E_b} p_e^E - (\tilde{d}_b - \tilde{w}_b)\}| \le F_l, \forall l \end{cases} \ge 1-\delta_L \quad (19)$$

$$\mathbb{P}_{\zeta_b^d,\zeta_b^\hbar}\begin{cases} |\sum_{b\in N} S_{l,b}^F\{\sum_{g\in G_b} p_g + \sum_{g\in G_b^c} r_g^\uparrow + \sum_{w\in W_b} p_w^W + \sum_{e\in E_b} p_e^E - (\tilde{d}_b - \tilde{w}_b)\}| \le F_l, \forall l \\ |\sum_{b\in N} S_{l,b}^F\{\sum_{g\in G_b} p_g - \sum_{g\in G_b^c} r_g^\downarrow + \sum_{w\in W_b} p_w^W + \sum_{e\in E_b} p_e^E - (\tilde{d}_b - \tilde{w}_b)\}| \le F_l, \forall l \end{cases} \ge 1-\delta_L \quad (20)$$

Under uncertain $\zeta_b^d, \zeta_b^\hbar$, the DC power flow constraints of the transmission line in the steady-state situation can also be expressed as JCC (19), which is intractable due to the product of the random variable and optimized variable, i.e., $\alpha_g \Delta p_L(\zeta)$. Therefore, a conservative relaxed model of the line power flow constraints [34] is adopted in (20), where the first row indicates the line flow constraint with dispatching the up reserve of AGC units, and the second row indicates the line flow constraint with dispatching the down reserve of AGC units. More details can be referred to [34].

In summary, due to the above nonconvex JCCs, the proposed JCED model is intractable. As discussed in Section I,

the Bonferroni inequality is used to transform the JCCs into a series of ICCs, which has to make *a priori* determination of the confidence level of the ICCs with the total number $Nw+2Ng+4Nl+2m+2$, which could be unfeasible for large systems. The traditional SAA [24] transforms the chance-constrained problem into an MILP model. However, a large number of sampling scenarios are required to ensure the SAA performance and badly exacerbate the complexity and computing burden of the optimization model. This issue will be discussed in detail in the next section.

## IV MSAA METHOD

In this section, a novel MSAA method is proposed to circumvent the shortcomings by leveraging the mixing inequalities and the aggregated mixing inequalities to formulate an LP model free of binary variables for a more efficient solution.

### A. General SAA transformation method

For a chance constraint $\mathbb{P}_\xi\{Ax \geq b(\xi)\} \geq 1-\delta$ with the random variable $\xi$ and the confidence level $1-\delta$, SAA uses $n$ sampling scenarios to represent the probability distribution $\mathbb{P}_\xi$, so that the chance constraint can be transformed into an MILP model shown in (21). More explanation about this formulation is stated in [24].

$$Ax = b + y \quad (21a)$$
$$y \geq w_i(1-z_i), \forall i \in [n] \quad (21b)$$
$$\sum_{i \in [n]} p_i z_i \leq \delta \quad (21c)$$
$$y \in \mathbb{R}_+, \ z_i \in \{0,1\}^n \quad (21d)$$

where $b = \min\{b(\xi^i), \forall i\}$; $w_i = b(\xi^i) - b$; $b(\xi^i)$ is the value of $b(\xi)$ under sampling scenario $i$, and $\sum_{i \in [n]} p_i = 1$ is satisfied.

As described above, the SAA equivalence is used to model the original chance constraint by making the number of scenarios in which $Ax \geq b(\xi^i)$ holds no less than $1-\delta$ of $n$. Obviously, only using a large number of sampling scenarios [n] can guarantee the accuracy of the SAA equivalence, which will yield a large-scale equivalent MILP problem.

### B. Basic Description of MSAA

An MSAA method is proposed to solve the proposed JCED model efficiently. Similar to SAA, the JCCs are transformed into the equivalent MILP model by using the equivalent probability distribution of sampling scenarios. Then, the binary indicator variables are relaxed into continuous variables. To reduce the resultant relaxation gap, the mixing inequalities and the aggregated mixing inequalities are applied to tighten the relaxed feasible region of the relaxed LP model. The details of these major procedures of the MSAA are described as follows.

*1) The transformation of JCCs in MSAA*
*a) Transformation of (13a) and (14)*

Since $H_W^{eq} + H_E^{eq} \geq 0, D_W^{eq} + D_E^{eq} \geq 0$ in (13a), $(\dot{a}),(\ddot{a})$ in (13a) can be directly equivalent to (22a) based on the SAA transformation idea in section A and [24], where both $(\dot{a})$ and $(\ddot{a})$ share the indicator variable $z_i^{sys}$ due to the original JCC formulation. Because the linear fitting coefficient depends on random disturbance $|\Delta p_L(\zeta)|$, $(\ddot{a})$ is difficult to transform. Considering that the larger the disturbance amplitude is, the higher the inertia requirements, the upper $\delta_F$ quantile of $|\Delta p_L(\zeta)|$ is used to calculate the linear fitting coefficients $\alpha_m^{\delta_F}, \beta_m^{\delta_F}$ in (22b).

$$\begin{cases} H_W^{eq}+H_E^{eq}\geq w_i^{rate}(1-z_i^{sys}), \ D_W^{eq}+D_E^{eq}\geq w_i^{ss}(1-z_i^{sys}), \forall i\in[n] \\ w_i^{rate}=\dfrac{|\Delta p_L(\zeta^i)|}{2\Delta \overline{f}_{rate}}-H_G^{eq}, \ w_i^{ss}=\dfrac{|-\Delta p_L(\zeta^i)|}{\Delta \overline{f}_{ss}}-(D_O+R_G^{-1}) \end{cases} \quad (22a)$$

$$H_W^{eq}+H_E^{eq}\geq \alpha_m^{\delta_F}-\beta_m^{\delta_F}(D_W^{eq}+D_E^{eq}), \forall m \quad (22b)$$

Similarly, the up-reserve constraints (14) for DIBRs are converted to (23), where $b_w^W = \min\{-\overline{\tilde{p}}_{i,w}^W\}$, $w_{i,w}^W = -\overline{\tilde{p}}_{i,w}^W - b_w^W$, $w_{i,w}^W \geq 0$. $z_i^W$ is the binary indicator variable to indicate if the up reserve constraint of DIBR is binding. $y_w^W \geq 0$ is the introduced auxiliary variable.

$$-p_w^W - (2H_w\Delta\overline{f}_{rate} + D_w\Delta\overline{f}_{max}) = b_w^W + y_w^W \quad (23a)$$
$$y_w^W \geq w_{i,w}^W(1-z_i^W) \quad (23b)$$
$$\sum_{i\in[n]} p_i z_i^W \leq \delta_{sys} \quad (23c)$$
$$H_w, D_w \in \mathbb{R}_+, \ z_i^W \in \{0,1\}^n \quad (23d)$$

*b) Transformation of SFR constraints for thermal units*

Since the coefficient of optimized $\alpha_g$ in (17) is the random variable $\Delta p_L(\zeta)$, the JCC in (17) is also difficult to transform. To solve this issue, (17) and (18) are replaced by the individual chance constraint (24), which limits the total amount of SFR reserves for all units. Hence, the total up/down reserve requirement $\Delta p_L^{\uparrow \delta_R/2}, \Delta p_L^{\downarrow \delta_R/2}$ can be solved by the quantile principle, and $\alpha_g$ can be optimized in (25) to achieve the optimal economic reserve allocation of each thermal unit, where $\Delta p_L^{\uparrow \delta_R/2}, \Delta p_L^{\downarrow \delta_R/2}$ denotes the upper/lower $\delta_R/2$ quantile of $\Delta p_L(\zeta)$. The value of $\delta_R$ is equal to $\delta_{SFR}$ in practice.

$$\mathbb{P}_{\Delta p_L(\zeta)}\{-\sum_{g\in G} r_g^\downarrow \leq \Delta p_L(\zeta) \leq \sum_{g\in G} r_g^\uparrow\} \geq 1-\delta_R \quad (24)$$

$$r_g^\uparrow \geq \alpha_g \Delta p_L^{\uparrow \delta_R/2}, -r_g^\downarrow \leq \alpha_g \Delta p_L^{\downarrow \delta_R/2}, \forall g \quad (25)$$

*c) Transformation of transmission line power flow constraints*

To simplify the notations, the transmission line power flow constraint is recast in (26a), and the definitions of the parameters are shown in (26b)-(26d).

$$A_l^* x \geq -F_l + e_l(\xi), \ -A_l^* x \geq -F_l - e_l(\xi), \ * \in \{\uparrow,\downarrow\} \quad (26a)$$

$$A_l^\uparrow x = \sum_{b\in N} S_{l,b}^F \{\sum_{g\in G_b} p_g + \sum_{g\in G_b^c} r_g^\uparrow + \sum_{w\in W_b} p_w^W + \sum_{e\in E_b} p_e^E\} \quad (26b)$$

$$A_l^\downarrow x = \sum_{b\in N} S_{l,b}^F \{\sum_{g\in G_b} p_g + \sum_{g\in G_b^c} r_g^\downarrow + \sum_{w\in W_b} p_w^W + \sum_{e\in E_b} p_e^E\} \quad (26c)$$

$$e_l(\xi) = \sum_{b\in N} S_{l,b}^F(\tilde{d}_b - \tilde{w}_b) \quad (26d)$$

The right sides of (26a) are denoted as $b_l^{low}(\xi) = -F_l + e_l(\xi)$ and $b_l^{up}(\xi) = -F_l - e_l(\xi)$. Then, the JCC in (20) can be transformed into (27), where $y_l^{*low}, y_l^{*up}$ are the introduced auxiliary variables, and $b_l^{low}, b_l^{up}, w_{l,i}^{up}, w_{l,i}^{low}$ are calculated in the previous section.

$$A_l^* x = b_l^{low} + y_l^{*low}, \ -A_l^* x = b_l^{up} + y_l^{*up}, \ * \in \{\uparrow,\downarrow\} \quad (27a)$$
$$y_l^{*low} \geq w_{l,i}^{low}(1-z_i^L), \ y_l^{*up} \geq w_{l,i}^{up}(1-z_i^L), \forall i \quad (27b)$$

$$\sum_{i\in[n]} p_i z_i^L \leq \delta_L \tag{27c}$$

*2) The Relaxed LP Model in MSAA*

Directly relaxing the binary variables into continuous variables in the range [0,1] is simple but inaccurate, since it may loosely enlarge the feasible region. To tighten the relaxed feasible region, the mixing inequalities and the aggregated mixing inequalities are introduced as follows. The mixing inequalities are mainly used for one-sided constraints, and the aggregated mixing inequalities are used for two-sided constraints [22].

*a) Mixing inequalities*

After directly relaxing the binary variables in (21), its feasible region $R_D$ can be expressed as (28). To tighten $R_D$, the mixing inequalities in (29) are involved, and the tightened feasible domain $R_T$ of the relaxed LP model is obtained as shown in (30). The relevant proof is stated in [35].

$$R_D = \{(y, z_i) \mid y \geq w_i(1-z_i),\ y \in \mathbb{R}_+, z_i \in [0,1], \forall i \in [n]\} \tag{28}$$

$$y + \sum_{s \in [\tau]} (w_{js} - w_{js+1}) z_{js} \geq w_{j1} \tag{29}$$

$$R_T = \{(y, z_i) \mid y \geq w_i(1-z_i),\ y + \sum_{s\in[\tau]}(w_{js}-w_{js+1})z_{js} \geq w_{j1},\\ y \in \mathbb{R}_+, z_i \in [0,1], \forall i \in [n]\} \tag{30}$$

where $[\tau]=\{j_1 \to \cdots \to j_\tau\}$ represents a rearranged subsequence in $[n]$ and satisfies $w_{j1} \geq w_{j2} \geq \cdots \geq w_{j\tau}$ and $w_{j\tau+1}=0$.

The corresponding mixing inequalities that are associated with (22), (23) and (26) can be respectively formulated in (31)-(33) as follows, where $\tau_{rate}, \tau_{ss}, \tau_W, \tau_l^{dn}, \tau_l^{up}$ are the respective $[\tau]$ associated with (22), (23) and (26).

$$H_W^{eq} + H_E^{eq} + \sum_{s\in[\tau_{rate}]} [w_{i,js}^{rate} - w_{i,js+1}^{rate}] z_{js}^{sys} \geq w_{i,j1}^{rate} \tag{31a}$$

$$D_W^{eq} + D_E^{eq} + \sum_{s\in[\tau_{ss}]} [w_{i,js}^{ss} - w_{i,js+1}^{ss}] z_{js}^{sys} \geq w_{i,j1}^{ss} \tag{31b}$$

$$y_w^W + \sum_{s\in[\tau_W]} [w_{i,js}^W - w_{i,js+1}^W] z_{js}^W \geq w_{i,j1}^W \tag{32}$$

$$y_l^{*low} + \sum_{s\in[\tau_l^{dn}]} [w_{l,js}^{low} - w_{l,js+1}^{low}] z_{js}^L \geq w_{l,j1}^{low},\ * \in \{\uparrow,\downarrow\} \tag{33a}$$

$$y_l^{*up} + \sum_{s\in[\tau_l^{up}]} [w_{l,js}^{up} - w_{l,js+1}^{up}] z_{js}^L \geq w_{l,j1}^{up},\ * \in \{\uparrow,\downarrow\} \tag{33b}$$

Taking (14) as an example, after obtaining the mixed-integer formulation (23) and relaxing the binary variables $z_{js}^W$ and adding the mixing inequalities (32), the feasible domain of $R_T$ form with regard to the original JCC (14) can be finally obtained.

*b) Aggregated mixing inequalities*

For two-sided constraints, [22] indicates that the aggregated mixing inequalities can also be added to make $R_T$ tighter. Thus, for constraint (26) with two sides limitations, the aggregated mixed inequality in (34) is also required besides constraint (33).

$$2y_c + \sum_{s\in\tau_R}(v_{l,rs}^{low}-v_{l,rs+1}^{low})z_{rs}^L + \sum_{s\in\tau_G}(v_{l,gs}^{up}-v_{l,gs+1}^{up})z_{gs}^L \geq v_{l,r1}^{low}+v_{l,g1}^{up} \tag{34}$$

where $y_c, v_{l,rs}^{low}, v_{l,gs}^{up}$ are the auxiliary variables, which are introduced in Appendix D.

*C. The Holistic Procedure of MSAA*

The holistic procedure of MSAA is summarized in Table I and illustrated in Fig. 5. The feasible domain after directly relaxing binary variables shown in Fig. 5(b) is inaccurate. Then, the mixing inequalities and the aggregated mixing inequalities are added to bound the feasible domain and make the feasible domain of the relaxed LP in Fig. 5(c) closer to the original. Hence, the original MILP problem is transformed into an LP problem without binary variables.

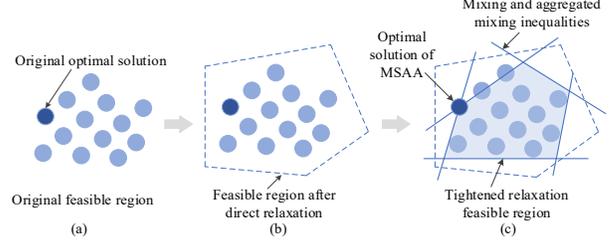

**Fig. 5.** Diagram of MSAA tightening the relaxation feasible domain.

TABLE I
THE HOLISTIC PROCEDURES OF MSAA

| | |
|---|---|
| 1 | //**for** JCC constraint (13a) Obtain the tightened LP model (22) and (31) with continuous variables $z_i^{sys}$ |
| 2 | //**for** JCC constraint (14) Obtain the tightened LP model (23) and (32) with continuous variables $z_i^W$ |
| 3 | //**for** JCC constraint (17) Obtain the equivalent LP model (25) |
| 4 | //**for** JCC constraint (20) Obtain the tightened LP model (27) and (33) with continuous variables $z_i^L$ |
| 5 | //Constraints in (6)-(18) except the above JCCS are reserved. |
| 6 | A tightened LP model is obtained, which is computationally cheaper than the MILP model resulted from SAA and more tightened than the direct 0-1 relaxation. |
| 7 | Use commercial solvers to solve this tightened LP model. |

## V. CASE STUDY

*A. Simulation settings*

In this section, the proposed JCED model and MSAA method are tested in the IEEE 39-bus and IEEE 118-bus systems with a resolution of 15 mins. The load and renewable energy data from the California ISO are applied in the simulation. The forecasting errors of loads and IBRs are assumed to follow a beta distribution [36], and the corresponding parameters are calculated based on historical data. The sampling scenarios (1000 sampling scenarios by default) are generated by Monte Carlo simulation (MCS). The following case studies were implemented on a computer with an Intel Core i7-11700 CPU and 16 GB RAM and solved with MATLAB and GUROBI 9.5.0, and the frequency dynamic response was verified in SIMULINK.

To test the effectiveness of the proposed model, three models are compared: 1) the JCED model with fixed inverter control parameters (Fix-JCED); 2) the individual chance-constrained ED (Up-ICED) [17] in which the constraints involving random variables are modeled as ICCs and only low-frequency related constraints are considered; and 3) the proposed JCED (Po-JCED).

In addition, the threshold values of the maximum ROCOF, the maximum frequency deviation and the steady-state frequency are set to 0.5 Hz/s, 0.5 Hz and 0.25 Hz, respectively. The significance level $\delta_F$ is set to 0 for enhanced frequency security from a conservative perspective, and the significance levels $\delta_{DIBR}, \delta_{SFR}, \delta_L$ are all set to 0.05. The regulation parameters for thermal units, DIBRs and ESs are detailed in [17].

The network and thermal unit parameters of the IEEE 39-bus and 118-bus systems are given by [37]. The cost coefficients of the reserve capacity and the redispatch power are set to 0.4 and 1.2 multiples of the base-point power cost coefficients, respectively. DIBRs with the capacities of 300 MW, 300 MW, 200 MW and 200 MW and four 25 MW/50 MWh ESs [38] are installed in the 39-bus system, and six 300 MW DIBRs and six 30 MW/30 MWh ESs are installed in the 118-bus system, where the charging/discharging efficiency of the ESs are set to 0.90/0.95.

*B. Verification of the Advantages of the Proposed JCED Model*

*1) Efficacy of considering both high- and low-related frequency constraints*

To indicate the significance of considering the inverter control parameter optimization and both high- and low-frequency related constraints, the three ED models listed above are simulated at the net load demand valley, e.g., 12 P.M. shown in Fig. 1. The scheduling results of the three cases are shown in TABLE II, where "*Average droop coefficient*" and "*Average virtual inertia*" represent the average values of the droop coefficients and virtual inertia of all DIBRs or ESs.

TABLE II
SCHEDULING RESULTS IN THE IEEE 39-BUS SYSTEM

| Models | | Fix-JCED | Up-JCED | Po-JCED |
|---|---|---|---|---|
| Total reserve of all thermal units (MW) | up | 538.18 | 496.02 | 538.18 |
| | down | 536.57 | 111.33 | 536.57 |
| Total reserve of all ESs (MW) | up | 8.00 | 20.00 | 20.00 |
| | down | 8.00 | 0.00 | 20.00 |
| Total up reserve of all DIBRs (MW) | | 230.86 | 264.59 | 302.55 |
| Average droop coefficient (*p.u.*) | DIBR | 4.00 | 8.79 | 8.41 |
| | ES | 8.00 | 10.00 | 10.00 |
| Average virtual inertia (*p.u.*) | DIBR | 2.00 | 2.98 | 3.40 |
| | ES | 4.00 | 5.00 | 5.00 |

TABLE II shows that the proposed Po-JCED model overwhelm other models in both over- and low-frequency dynamic performance. Specifically, the virtual inertia and the droop coefficients of Fix-JCED are lower than those of UP-ICED and Po-JCED, because the latter two both optimized the inverter control parameters. Therefore, the maximum frequency deviation of Fix-JCED can be higher than that of UP-ICED and Po-JCED. This is confirmed in Fig. 6, which shows the frequency response values under the various disturbances in the 39-bus system. Figs. 6b and 6d compared the dynamic results with the disturbances of 20% and 25% of the net load, which cause high- and low-frequency events, respectively, the results of Fix-JCED are less stable than Po-JCED and exceed the threshold value of the maximum frequency deviation in Fig. 6d.

Furthermore, TABLE II confirms that UP-ICED has smaller down reserves than Po-JCED due to the shortcomings of Up-ICED pointed out in Section I, which could result in a larger steady-state frequency deviation, as shown in Figs. 6a and 6b. This larger deviation might cause a higher operational risk and even trigger protection devices. In the low-frequency case shown in Figs. 6c and 6d, Po-JCED and Up-ICED perform similarly. However, since JCCs can avoid the problems that may be caused by ICC, as mentioned in Section III, in the cases shown in Fig. 6d, Po-JCED can perform better than Up-ICED, which will be analyzed later.

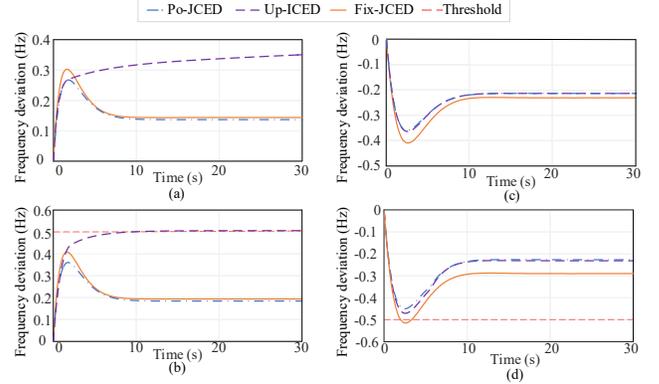

**Fig. 6.** Frequency dynamic response under various disturbances in the IEEE 39-bus system. In the case of net load reduction, (a) disturbance 1 and (b) disturbance 2 are 15% and 20% of the net load, respectively; in the case of net load increase, (c) disturbance 3 and (d) disturbance 4 are 20% and 25% of the net load, respectively.

*2) Verification of the Efficacy of Adopting the JCC Formulation*

Tables III and IV show the probabilities of reserve deficiency of Up-ICED and Po-JCED in the 39-bus and 118-bus systems, which are evaluated by using 10,000 testing scenarios generated by MCS. It can be indicated that Po-JCED provides scheduling decisions with a higher reliability level of the up reserve of DIBR, because the JCCs are more secure to model the reserve constraints than ICCs [18]. In addition, since the common feasible region for all ICCs is smaller than the feasible region of the corresponding JCC [39], Po-JCED can provide more secure dispatch strategy for SFR reserve and transmission line power flow than Up-ICED.

TABLE III
PROBABILITY OF RESERVE DEFICIENCY IN THE IEEE 39-BUS SYSTEM IN 10000 TESTING SCENARIOS

| Model | Up reserve of DIBR | SFR reserve |
|---|---|---|
| Up-ICED | 12.69% | 10.31% |
| Po-JCED | 3.08% | 2.81% |

TABLE IV
PROBABILITY OF RESERVE DEFICIENCY IN THE IEEE 118-BUS SYSTEM IN 10000 TESTING SCENARIOS

| Model | Up reserve of DIBR | SFR reserve | Line power flow |
|---|---|---|---|
| Up-ICED | 17.89% | 9.49% | 4.02% |
| Po-JCED | 3.54% | 4.35% | 2.35% |

Furthermore, it is worth noting that due to higher insufficient reserve probabilities, the scheduling results of Up-ICED could increase the risk of frequency violations. For example, with a high-risk testing scenario with a 27.5% disturbance, the frequency change curve in the 39-bus system is shown in Fig. 7. Under this disturbance, the frequency support performance of DIBRs cannot be fully achieved in Up-ICED due to the restricted reserve of unit 2, so the maximum frequency deviation exceeds its limit. In contrast, Po-JCED can better recover the frequency.

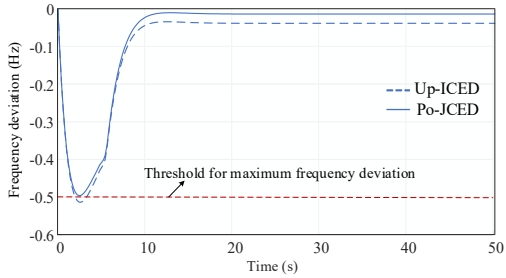

Fig. 7. Frequency dynamic curve under high-risk disturbance in IEEE 39-bus system.

In addition, the costs of Up-ICED and Po-JCED in the 39-bus and 118-bus systems are compared in TABLE V, where the "*objective cost*" refers to the cost in (6), the "*ex post cost*" refers to the expectations of load shedding and the IBR spillage cost when the reserves are insufficient in testing scenarios, the coefficient of "*ex post cost*" is set at $5000/MWh according to[40] and the "*total cost*" is the sum of the two costs. Compared to Up-ICED, Po-JCED has higher *objective costs* but lower *ex post costs* and lower *total costs*, because Po-JCED considers both up and down overlimit issues of the system frequency and adopts JCC formulations. With more dispatched reserve capacity, the system reliability is improved, and the *total cost* is reduced.

TABLE V
COST OF THE DISPATCH MODEL IN THE IEEE 39-BUS SYSTEM AND IEEE 118-BUS SYSTEM WITH 10,000 TESTING SCENARIOS (UNIT: $)

| System | Model | Objective cost | Ex post cost | Total cost |
|---|---|---|---|---|
| 39-bus | Up-ICED | 22034.19 | 11187.33 | 33221.52 |
| | Po-JCED | 24596.33 | 1979.61 | 26575.94 |
| 118-bus | Up-ICED | 32977.39 | 10092.19 | 43069.58 |
| | Po-JCED | 37055.51 | 2852.53 | 39908.04 |

*3) Economic benefits by optimizing the AGC factors*

The objective cost reduction ratio of the Po-JCED model with adjustable AGC factors in the 39-bus system is shown in TABLE VI, where "*disturbance level*" refers to the proportion of the power disturbance to the predicted net load, and "*cost reduction ratio*" represents the ratio of the objective cost reduction of Po-JCED using adjustable AGC factors over Po-JCED with fixed AGC factors. The results indicate that with the increase of the disturbance level, the objective cost is significantly reduced by optimizing the AGC allocation factors.

TABLE VI
COST REDUCTION RATIO DUE TO ADJUSTABLE AGC FACTOR IN IEEE 39-BUS SYSTEM WITH 1000 SAMPLING SCENARIOS

| Disturbance level/% | 0 | 5 | 10 | 15 | 20 |
|---|---|---|---|---|---|
| Cost reduction ratio/% | 0 | 3.49 | 5.36 | 5.79 | 6.34 |

*C. Computational Performance*

TABLE VII
OBJECTIVE COST AND COMPUTATIONAL TIME IN IEEE 39-BUS AND IEEE 118-BUS SYSTEMS WITH 1000 SAMPLING SCENARIOS

| System | Solution | Objective cost/$ | Computational time/s |
|---|---|---|---|
| 39-bus | SAA | 24596.33 | 7.61 |
| | MSAA | 24471.37 | 0.48 |
| 118-bus | SAA | 37055.51 | 8.52 |
| | MSAA | 36618.81 | 1.77 |

The computational time of the original SAA method and the proposed MSAA method for solving Po-JCED with 1,000 sampling scenarios in the 39-bus and 118-bus systems are shown in TABLE VII. Encouragingly, compared with the SAA method, the MSAA method reduces the computational time by 93.69% with 0.51% accuracy loss in the IEEE 39-bus system, and the MSAA reduced the computation time by 79.23% with 1.18% accuracy loss in the IEEE 118-bus system.

To further demonstrate the efficiency of the MSAA method in large-scale systems, the computational time curves of MSAA and SAA in the IEEE 118-bus system are shown in Fig. 8, which illustrates that the proposed MSAA method can save more computational time as the number of scenarios increases. For example, with 3,000 sampling scenarios, the SAA calculation time is 140.82 seconds, while the MSAA only costs 4.70 seconds, a 96.66% reduction. Moreover, the accuracy loss of the MSAA is within an acceptable range, e.g., the cost error is in the range of 0.54%-1.77% in the case of 500-3,000 scenarios, where the cost error is the ratio of the cost difference between SAA and MSAA over the SAA cost[2].

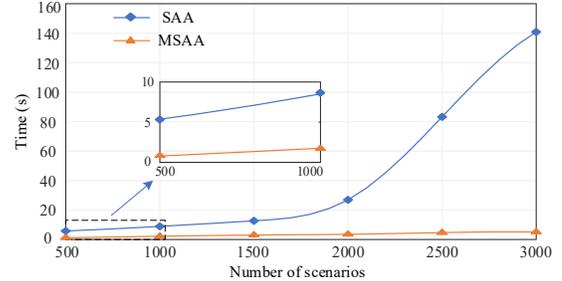

Fig. 8. Computational time of SAA and MSAA as the number of sampling scenarios increases in the IEEE 118-bus system.

## VI. CONCLUSIONS

In this paper, a JCED model involving the joint optimization of frequency-related controller parameters and up/down regulation reserves is proposed. In this model, the JCC formulations are adopted, and the AGC allocation factors of thermal units, virtual inertia and droop control coefficients of DIBRs and ESs are optimized, which significantly improve the operation economy and power system security. Moreover, an MSAA method is proposed to solve JCED efficiently. Simulation results show that the proposed JCED model significantly increases the operational frequency security level compared with Up-ICED, and the operational cost is reduced by optimizing AGC factors. In addition, the proposed MSAA method significantly reduces the computational burden compared to traditional SAA methods.

It is worth mentioning that the proposed JCED model will play a more significant role in the power systems with high penetration of renewable energy, and the MSAA method also has great potential to efficiently solve other chance-constrained dispatch models for large-scale systems. Future research can be conducted involving optimizing the number and locations of DIBRs and ESs, as well as the consideration of discrete droop coefficients in inverter control.

## APPENDIX A.

### AGGREGATION PARAMETERS OF THE SYSTEM

In the power system with thermal units, DIBRs and ESs, an aggregated system swing equation [4] can be expressed as (A1). Then, the equivalent model (2) of the system frequency response can be obtained by substituting (1) into (A1).

---

[2] Since there is no direct and accurate method to deal with the JCCs, the results of the SAA method are used as the optimality benchmark, which will be accurate with enough sampling scenarios[41].

$$\Delta p_W^{eq}+\Delta p_E^{eq}+\Delta p_G^{eq}-\Delta p_L = 2H_G^{eq}\frac{d\Delta f}{dt}+D_O\Delta f \quad (A1)$$

where the aggregated parameters $H_G^{eq}, H_W^{eq}, H_E^{eq}, D_W^{eq}, D_E^{eq}, R_G, F_H, T_R$ in the equation are calculated [27] as follows.

$$H_G^{eq}=\sum_{g\in G}H_g\frac{\bar{p}_g}{p_{sys}},\ \frac{1}{R_G}=\sum_{g\in G}\frac{1}{R_g}\frac{\bar{p}_g}{p_{sys}} \quad (A2a)$$

$$H_W^{eq}=\sum_{j\in W}H_j\frac{p_j^{cap}}{p_{sys}},D_W^{eq}=\sum_{j\in W}D_j\frac{p_j^{cap}}{p_{sys}} \quad (A2b)$$

$$H_E^{eq}=\sum_{e\in W}H_e\frac{\bar{p}_e^E}{p_{sys}},D_E^{eq}=\sum_{j\in W}D_e\frac{\bar{p}_e^E}{p_{sys}} \quad (A2c)$$

$$\lambda_g=\frac{1}{R_g}\times\frac{\bar{p}_g}{p_{sys}}\bigg/\sum_{g\in G}\frac{1}{R_g}\times\frac{\bar{p}_g}{p_{sys}} \quad (A2d)$$

$$F_H=\sum_{g\in G}\lambda_g F_g^H, T_R=\sum_{g\in G}\lambda_g T_g^R \quad (A2e)$$

where $\Delta p_W^{eq},\Delta p_E^{eq}$ represent the output variation of the aggregated DIBR and ES, respectively; $p_{sys}$ represent the total generation capacity of the system; $H_g, F_g^H, T_g^R$ represent the inertia, the fraction of power generated by the high-pressure turbine and the reheat time of thermal unit $g$, respectively; $p_j^{cap}$ represents the generation capacity of DIBR $j$; and $\lambda_g$ represents the defined nonlinear gain[27].

## APPENDIX B.
### DERIVATION OF FREQUENCY-RELATED CONSTRAINTS

*(1) Frequency-related indices*

As is widely used by [4]-[6], the maximum RoCoF, the maximum frequency deviation $\Delta f_{max}$ and the steady-state frequency deviation $\Delta f_{ss}$ can be expressed as follows.

$$RoCoF = \frac{-\Delta p_L}{2(H_G^{eq}+H_W^{eq}+H_E^{eq})} \quad (B1)$$

$$\Delta f_{max}=\frac{-\Delta p_L}{D_{sys}+R_G^{-1}}\times(1+\sqrt{T_R^2\omega_n^2-2\zeta\omega_n T_R+1}\times e^{-\zeta\omega_n t_{max}})=\frac{-\Delta p_L}{F_{un}} \quad (B2)$$

$$\Delta f_{ss}=\frac{-\Delta p_L}{D_O+D_W^{eq}+D_E^{eq}+R_G^{-1}} \quad (B3)$$

where $\zeta,\omega_n,t_{max}$ can be calculated by the method in [6], as long as the equivalent system inertia $H_{sys}=H_G^{eq}+H_W^{eq}+H_E^{eq}$ and equivalent system damping $D_{sys}=D_O+D_W^{eq}+D_E^{eq}$ are substituted.

*(2) The linear maximum frequency deviation constraint*

The maximum allowed $\Delta f_{max}$ can be expressed as (B4). As the unit commitment is fixed in the ED, $H_G^{eq}, R_G^{-1}$ are regarded as constants. Thus, $\Delta f_{max}$ is a nonlinear function of $H_W^{eq}, H_E^{eq}, D_W^{eq}, D_E^{eq}$, which is intractable in the optimization model.

$$|\Delta f_{max}|=|\frac{-\Delta p_L}{F_{un}(H_W^{eq},H_E^{eq},D_W^{eq},D_E^{eq})}|\leq\Delta\bar{f}_{max} \quad (B4)$$

Furthermore, to make nonlinear (B4) tractable, we study the $H^I-D^I$ safe boundary curve [9] of $|\Delta f_{max}|$, where $H^I=H_W^{eq}+H_E^{eq}$, $D^I=D_W^{eq}+D_E^{eq}$. When the point $(H^I,D^I)$ is in the region above the boundary, $|\Delta f_{max}|$ is within the threshold value. Hence, incorporating the boundary constraints into ED can guarantee (B4). To obtain the linear boundary expression, the piecewise linearization method shown in (B5) is utilized to establish a linear relationship between $H^I$ and $D^I$, i.e., $h(D^I)=\max_{1\leq m\leq M}\{\alpha_m-\beta_m D^I\}$. Thus, the nonlinear (B4) can be transformed into a linear formulation (B6).

$$\min_{\alpha_m,\beta_m}\sum_{k\in\Psi}(\max_{1\leq m\leq M}\{\alpha_m-\beta_m D_k^{eq}\}-H_k^{eq})^2$$
$$s.t.\ \alpha_m-\beta_m D_k^{eq}\geq H_k^{eq},\ \forall k\in\Phi_m \quad (B5)$$

$$H_W^{eq}+H_E^{eq}\geq\alpha_m-\beta_m(D_W^{eq}+D_E^{eq}),\forall m \quad (B6)$$

where $(H_k^{eq},D_k^{eq})$ represents a point in the original boundary, $\Phi_m$ represents a set of points in the $m^{th}$ piece with $M$ pieces, $\Psi=\{\Phi_1...\Phi_M\}$, and $\alpha_m,\beta_m$ represent the corresponding fitting coefficients.

*(3) Relaxed PFR constraints of DIBRs and ESs*

The right-hand side of constraint (4) can be relaxed to constraint (B7) by using the well-known triangle inequality, and then the more conservative constraint (5) can be readily obtained.

$$2H_k\frac{d\Delta f}{dt}+D_k\Delta f|\leq 2H_k|\frac{d\Delta f}{dt}|+D_k|\Delta f|\leq 2H_k\Delta\bar{f}_{rate}+D_k\Delta\bar{f}_{max} \quad (B7)$$

## APPENDIX C.
### APPROXIMATION OF REDISPATCH COSTS

To overcome the nonconvexity of the redispatch cost in (6), the following linear approximation formulation (C1) is adopted. The approximation is valid, because *a)* the approximation does not affect the constraints, so it does not affect the feasible region of this problem; *b)* the $\delta_{SFR}$ of the corresponding JCC (17) is generally small (e.g., 0.05 in the case study), so (C1) has little influence on the operating economy; and *c)* the JCED model can be transformed into a computationally efficient LP model after using the (C1) approximation and MSAA processing. Finally, after solving the optimized model with approximation (C1), the original model in (6) is used to update the redispatch cost.

$$\sum_{g\in G}\mathbb{E}_{\Delta p_L(\zeta)}c_g^r(\min\{r_g^+(\zeta),r_g^\uparrow\}+\min\{r_g^-(\zeta),r_g^\downarrow\})$$
$$\approx\sum_{g\in G}\mathbb{E}_{\Delta p_L(\zeta)}(c_g^r|\alpha_g\Delta p_L(\zeta)|) \quad (C1)$$

## APPENDIX D.
### AGGREGATED MIXING INEQUALITIES FOR TRANSMISSION LINES

The derivation of the aggregated mixing inequality is detailed in [20]. In this paper, the aggregated mixing inequality is applied directly to transform the two-sided constrained JCC of transmission lines. First, $2F_l$ is added to both sides of (26), as shown in (B1).

$$2F_l+A_l^*x\geq F_l+e_l(\xi),\ *\in\{\uparrow,\downarrow\}$$
$$2F_l-A_l^*x\geq F_l-e_l(\xi),\ *\in\{\uparrow,\downarrow\} \quad (B1)$$

Letting $y_c=2F_l, y_a=A_l^*x$, $v_{l,i}^{low}=F_l+e_l(\xi^i)$, $v_{l,i}^{up}=F_l-e_l(\xi)$, and $v_{l,i}^{low}\geq v_{l,i}^{up}\geq 0$, the auxiliary variable $u_a \geq y_a \geq 0$ is applied, and $u_a \geq \max\{v_{l,i}^{low}, i \in [n]\}$. Then, (B1) can be reformulated as (B2), where (B2b) is used to ensure that $y_c - y_a \geq v_{l,i}^{up}$ when $z_i^L = 0$; otherwise, $y_c - y_a + u_a \geq 0$ when $z_i^L = 1$. Meanwhile, (B2a) is used to maintain $y_c + y_a \geq 0$.

$$y_c + y_a + v_{l,i}^{low} z_i^L \geq v_{l,i}^{low} \quad \text{(B2a)}$$

$$y_c - y_a + (v_{l,i}^{up} + u_a) z_i^L \geq v_{l,i}^{up} \quad \text{(B2b)}$$

$$u_a \geq y_a \geq 0, \ y_c \geq 0, \ z_i^L \in \{0,1\} \quad \text{(B2c)}$$

Letting $y_1 = y_c + y_a, y_2 = y_c - y_a + u_a$, (B2) can be reorganized as (B3). Based on the literature [20], given any subsequence $\Theta = \{i_1 \rightarrow \cdots \rightarrow i_\theta\}$ in [n], the aggregated mixing inequality to be added is (B4), where $z_{i\theta}^L$ represents the indicator variable for the last scenario in $\Theta$, and $\tau_R, \tau_G$ are the 1-mixing subsequence and the 2-mixing subsequence of $\Theta$, respectively. In addition, $z_{g\tau_G}^L = z_{i\theta}^L$, $(v_{l,g\tau_G}^{up} - v_{l,g\tau_G+1}^{up}) z_{\tau_G}^L = (v_{l,i}^{up} + u_a) z_{\tau_G}^L$. The above conditions are combined to transform (B4) into (B5), which is equation (34).

$$y_1 + v_{l,i}^{low} z_i^L \geq v_{l,i}^{low} \quad \text{(B3a)}$$

$$y_2 + (v_{l,i}^{up} + u_a) z_i^L \geq (v_{l,i}^{up} + u_a) \quad \text{(B3b)}$$

$$y_1 + y_2 \geq u_a, \ y_1 \geq 0, \ y_2 \geq 0 \quad \text{(B3c)}$$

$$z_i^L \in \{0,1\}^n \quad \text{(B3d)}$$

$$y_1 + y_2 + \sum_{s \in \tau_R}(v_{l,rs}^{low} - v_{l,rs+1}^{low})z_{rs}^L + \sum_{s \in \tau_G}(v_{l,gs}^{up} - v_{l,gs+1}^{up})z_{gs}^L - u_a z_{i\theta}^L \geq v_{l,r1}^{low} + v_{l,g1}^{up} + u_a \quad \text{(B4)}$$

$$2y_c + \sum_{s \in \tau_R}(v_{l,rs}^{low} - v_{l,rs+1}^{low})z_{rs}^L + \sum_{s \in \tau_G}(v_{l,gs}^{up} - v_{l,gs+1}^{up})z_{gs}^L \geq v_{l,r1}^{low} + v_{l,g1}^{up} \quad \text{(B5)}$$